\documentclass[12pt]{article}
\usepackage{bbm}
\usepackage{amsmath,amssymb,amsfonts,epsfig,graphicx,euscript}
\usepackage[numbers,sort&compress]{natbib}
\usepackage[margin=1.0in]{geometry}
\usepackage[bookmarks,bookmarksnumbered,linktocpage,pdfstartview=FitH]{hyperref}
\hypersetup{colorlinks,%
citecolor=red,%
filecolor=blue,%
linkcolor=blue,%
urlcolor=blue,%
pdftex}
\usepackage{color}
\usepackage{xcolor}
\usepackage{amsmath}
\usepackage{amsfonts}
\usepackage{graphicx}
\usepackage{caption}
\usepackage{subcaption}
\usepackage{float}
\usepackage[all]{hypcap}
\numberwithin{equation}{section}
\usepackage{calc}
\usepackage{physics}
\usepackage[titletoc]{appendix}
\allowdisplaybreaks
\newlength{\spacer}
\newsavebox{\mybox}

\newcommand{\bse}{\begin{subequations}}
\newcommand{\ese}{\end{subequations}}
\newcommand{\be}{\begin{equation}}
\newcommand{\ee}{\end{equation}}
\newcommand{\bea}{\begin{eqnarray}}
\newcommand{\eea}{\end{eqnarray}}
\newcommand{\ba}{\begin{array}}
\newcommand{\ea}{\end{array}}

\newcommand{\mch}{\mathcal{H}}
\newcommand{\hmch}{\hat{\mathcal{H}}}

\newcommand
\numberthis{\addtocounter{equation}{1}\tag{\theequation}}

\begin{document}

\renewcommand*{\thefootnote}{\fnsymbol{footnote}} 

\begin{center}

{ \large{\textbf{Fermion Number 1/2 of Sphalerons and Spectral Mirror Symmetry}}} 
\vspace*{0.5cm}
\begin{center}
{\bf M. Mehraeen\footnote{man$_{-}$mehrloco@yahoo.com} and S. S. Gousheh\footnote{ss-gousheh@sbu.ac.ir}}\\%
\vspace*{0.5cm}
{\it {Department of Physics, Shahid Beheshti University, G.C., Evin, Tehran 19839, Iran\\}}  
\vspace*{0.5cm}
\end{center}
\end{center}

\renewcommand*{\thefootnote}{\arabic{footnote}} 
\setcounter{footnote}{0}

\begin{center}
\today

\bigskip
\textbf{Abstract}
\end{center}

We present a rederivation of the baryon and lepton numbers $\frac{1}{2}$ of the $SU(2)_L$ S sphaleron of the standard electroweak theory based on spectral mirror symmetry. We explore the properties of a fermionic Hamiltonian under discrete transformations along a noncontractible loop of field configurations that passes through the sphaleron and whose endpoints are the vacuum. As is well known, CP transformation is not a symmetry of the system anywhere on the loop, except at the endpoints. By augmenting CP with a chirality transformation, we observe that the Dirac Hamiltonian is odd under the new transformation precisely at the sphaleron, and this ensures the mirror symmetry of the spectrum, including the continua. As a consistency check, we show that the fermionic zero mode presented by Ringwald in the sphaleron background is invariant under the new transformation. The spectral mirror symmetry which we establish here, together with the presence of the zero mode, are the two necessary conditions whence the fermion number $\frac{1}{2}$ of the sphaleron can be inferred using the reasoning presented by Jackiw and Rebbi or, equivalently, using the spectral deficiency $\frac{1}{2}$ of the Dirac sea. The relevance of this analysis to other solutions is also discussed.

\newpage


\section{Introduction}\label{Introduction}
In their seminal paper on the subject of charge fractionalization, Jackiw and Rebbi studied the Dirac equation in classical bosonic backgrounds for a number of field theories  \cite{jackiw76}. Their key discovery was the existence of states with half-integer fermion numbers in theories where all the fundamental fields have integer fermion numbers. As was pointed out in \cite{jackiw76, jackiw77, goldwilczek81, mackenzie84}, in order for a bosonic configuration to have half-integer fermion numbers, the following two conditions must be simultaneously satisfied:
\begin{enumerate}
\item The existence of a normalizable fermionic zero mode precisely at the
bosonic configuration;
\item Mirror symmetry of the entire fermionic spectrum, consisting of the bound and continuum states, at the bosonic configuration, or equivalently, the fermion number conjugation invariance of the system.
\end{enumerate}

Since then, charge fractionalization has been widely studied and has found many applications in different areas, such as particle physics \cite{jackiw77, goldwilczek81, mackenzie84, witten79, macwilczek84, gousheh94,  charmchi14}, condensed matter physics \cite{su79,su81, semen86}, polymer physics \cite{rice82, jacksemen83, heeger88}, quantum wires \cite{steinberg08} and topological insulators \cite{maciejko15, stern16}.

One class of solutions that can be found in certain field theories are sphalerons, which are saddle-point solutions in field configuration space \cite{manton83, manton04}. 
An important member of this class of solutions is the `S' sphaleron \cite{klink03} of the standard electroweak theory. Its importance is due to the role that it is believed to play in the early Universe, including the generation of the matter-antimatter asymmetry of the Universe \cite{klink84, rubakov96, shiva18}. Following the discovery of this solution in hadronic models \cite{dhn74,boguta83}, it was rediscovered \cite{manton83} in $SU(2)_L$ theory and its properties and implications for cosmology were detailed in \cite{klink84}.

In 1974 Dashen et.\ al.\ not only constructed a sphaleron solution as an extended model of hadrons, but also presented a framework for calculating the bound state energies of fermions which couple to the $SU(2)$ gauge field component of the sphalerons \cite{dhn74}. The coupling of the fermions to the Higgs component was represented as an explicit fermionic mass term. 	
Based on their work, the author of \cite{nohl75} showed that in the classical $SU(2)$ gauge field of the sphaleron, a fermion has a single bound-state solution with zero energy. In the standard electroweak theory, a single normalizable zero energy solution of the Dirac equation in the sphaleron background was shown to exist for massless fermions in \cite{boguta85}. Shortly after, this result was extended by Ringwald to the case of massive fermions \cite{ringwald88}.

Later on, in the level-crossing picture for the $SU(2)_L$ theory, the change in the bound state energy of fermions coupled to the bosonic fields of the noncontractible loop (NCL) was numerically determined \cite{kunz93}. There, it was shown that as one traverses the NCL passing through the sphaleron, a single negative eigenvalue of the Dirac Hamiltonian arises from the Dirac sea, crosses the zero energy level precisely at the sphaleron and enters the positive energy continuum as one returns to the vacuum configuration. 
The numerical study of \cite{kunz93} thus reconfirmed the existence of a zero energy bound state in the sphaleron background.


It is well known that the baryon and lepton numbers of the S sphaleron are $\frac{1}{2}$ \cite{klink84}. This can be seen by using the chiral anomaly and integrating the temporal component of the Chern-Simons current over one-point compactified 3-space and obtaining the resultant Chern-Simons charge for the sphaleron configuration, which is half-integer due to $SO(3)$ and reflection symmetries in the bosonic sector \cite{manton04}. However, the use of the one-point compactification scheme corresponds to a gauge that breaks the reflection symmetry of the bosonic fields of the NCL about the sphaleron \cite{klink84, manton04}. On the other hand, the spectral flow for the Dirac Hamiltonian along the NCL and symmetries of the fermionic sector are independent of the gauge, and constitute the core of this paper.

The spectra of fermions coupled to the $SU(2)_L$ gauge-Higgs fields of the NCL passing through the S sphaleron have been studied in great detail over the past three decades by various authors. In doing so, many of the symmetries of the spectra have been pointed out and explored \cite{klink84, yaffe89, kunz93, kunz94, nolte96, manton04}. However, a symmetry that seems to have not been fully elucidated hitherto in the literature is a certain conjugation symmetry precisely at the sphaleron. This manifests itself as the mirror symmetry of the fermionic spectrum about $E=0$. Following the path of the bound state as the NCL is traversed, this symmetry can be obviously seen to exist for the bound state precisely at the sphaleron, where it crosses $E=0$. Now, the question is whether the entire fermionic spectrum, including the continuum states, has mirror symmetry at the sphaleron. As we shall show, this is indeed the case, and this has an important implication which brings us to the subject of this paper. 

The main goal of this paper is to present a rederivation of the half-integer fermion numbers of $SU(2)_L$ S sphalerons by adopting an approach that is based on discrete symmetries. To do this, we explicitly construct the transformation operator, which consists of the chiral and CP transformations, under which the Dirac Hamiltonian at the sphaleron is odd. Hence we show that the entire spectrum of the Dirac Hamiltonian has mirror symmetry in the presence of the sphaleron. We then use the results presented by Jackiw and Rebbi \cite{jackiw76} to argue that the presence of the zero mode mandates half-integer fermion numbers for the sphaleron. 

We should mention that the relation between the results of \cite{jackiw76} and sphalerons had been hinted at in works such as \cite{klink84, klink84b, boguta85, klink03}. However, as mentioned before, a necessary condition to make such a connection is the mirror symmetry of the entire spectrum, which we establish here.  Furthermore, whereas some works have based their arguments on a fermionic zero mode in the limit of vanishing fermion mass \cite{klink84b, boguta85}, in this work we have used the Ansatz given by \cite{ringwald88}, which is an extension to massive fermions within the standard electroweak theory. In this case, the Higgs component of the sphaleron plays a nontrivial and essential role, which is beyond an explicit mass term. 


The outline of this paper is as follows. In Section \ref{SU2Sphaleron}
, we briefly review the bosonic sector of the standard electroweak theory and the sphaleron Ansatz of $SU(2)_L$ Yang-Mills-Higgs theory in the limit of vanishing weak mixing angle. In Section \ref{CP Transformation along Noncontractible Loop}, we analyze the behavior of the Dirac Hamiltonian operator under a CP transformation for all configurations along the NCL. Then, we augment CP to arrive at a suitable choice of conjugation operator which reveals the spectral mirror symmetry at the sphaleron. Then, in Section \ref{The Zero Mode}, we perform the symmetry transformation on the zero mode given by Ringwald \cite{ringwald88} in the sphaleron background. In Section \ref{Summary and Discussion}, we summarize our results and present an outlook.

\section{$SU(2)_L$ Sphaleron}\label{SU2Sphaleron}

Consider the bosonic sector of the well-established electroweak Lagrangian
\be
\mathcal{L}= -\frac{1}{4}G_{\mu\nu}^a G^{\mu\nu} _a  -\frac{1}{4}F_{\mu\nu}F^{\mu\nu} + \left(D_{\mu} \Phi)^{\dagger}(D^{\mu}\Phi\right) - \lambda\left(\Phi^{\dagger} \Phi - \eta^2 \right)^2,
\ee
where the $U(1)$ field strength tensor is given by
\be
F_{\mu\nu} = \partial_{\mu}A_{\nu} - \partial_{\nu}A_{\mu},
\ee
the $SU(2)$ field strength tensor is given by
\be
G_{\mu\nu}^a = \partial_{\mu}B_{\nu}^a - \partial_{\nu}B_{\mu}^a + g\epsilon^{abc}B_{\mu}^b B_{\nu}^c,
\ee
and the covariant derivative of the Higgs field is
\be
D_{\mu}\Phi = \left(\partial_{\mu} - ig\frac{\tau^a}{2}B_{\mu}^a -ig^{\prime}YA_{\mu}\right)\Phi.
\ee
\if
The non-vanishing vacuum expectation value (VEV) of the Higgs field
\be
\left\langle \Phi \right\rangle = \eta
\begin{pmatrix}
0\\
1\\
\end{pmatrix}
\ee
spontaneously breaks the gauge symmetry such that
\be
SU(2)_L \times U(1)_Y \xrightarrow{\text{SSB}} U(1)_{EM}.
\ee
This ensures that the Maxwell field remains massless, while the masses of the Higgs and remaining gauge bosons are given by
\be
M_W=\frac{1}{\sqrt{2}}g\eta ,\;\;\;\;\;M_Z=\frac{1}{\sqrt{2}}\sqrt{g^2 + g^{\prime ^2}} \eta ,\;\;\;\;\;M_H=2\eta \sqrt{\lambda}.
\ee
Finally, the weak mixing angle $\theta_w$ and electric charge $e$ are determined by
\be
\tan\theta_w= \frac{g^{\prime}}{g},\;\;\;\;\;e=g \sin\theta_w.
\ee
\fi
In the limit of vanishing mixing angle, the $U(1)$ field decouples and this allows for a spherically symmetric Ansatz for the gauge and Higgs fields of the NCL. To this end, consider the following map:
\be
U: S^1 \wedge S^2 \sim S^3 \rightarrow SU(2),\;\;\;\;\;\;\left(\mu,\theta,\phi \right)\mapsto U\left(\mu,\theta,\phi\right),
\ee
where $\wedge$ is the \textit{smash} product\footnote{For a definition, see \cite{klink03}.} and $\mu$ is the loop parameter. A suitable representation is \cite{manton83,klink03}
\be\label{U}
U\left(\mu,\theta,\phi\right)=-iy^1 \tau_1 -iy^2 \tau_2 -iy^3\tau_3 + y^4 \mathbbm{1},
\ee
where
\be
\begin{pmatrix}
y^1\\
y^2\\
y^3\\
y^4\\
\end{pmatrix}=
\begin{pmatrix}
-\sin\mu\sin\theta\sin\phi\\
-\sin\mu\sin\theta\cos\phi\\
\sin\mu\cos\mu\left(\cos\theta-1\right)\\
\cos^2\mu + \sin^2\mu\cos\theta\\
\end{pmatrix},
\ee
and $\tau^i$ are chosen to be the usual Pauli matrices while $\tau^i /2$ are the generators in weak isospace. Using the above map, the Ansatz\footnote{It can be shown that, even when the Ansatz is not manifestly spherically symmetric, it can always be transformed to one that is \cite{klink90,witten77}.} for the static gauge and Higgs fields of the $SU(2)_L$ sphaleron barrier becomes \cite{manton83}
\be
\begin{split}\label{hedgehog Ansatz}
B\left(\mu,r,\theta,\phi\right)&=-\frac{f\left(r\right)}{g} dU\left(\mu,\theta,\phi\right)U^{-1}\left(\mu,\theta,\phi\right),\\
\Phi\left(\mu,r,\theta,\phi\right)&= \eta h\left(r\right)U\left(\mu,\theta,\phi\right)
\begin{pmatrix}
0\\
1\\
\end{pmatrix}
+\eta\left[1-h\left(r\right)\right]
\begin{pmatrix}
0\\
e^{-i\mu}\cos\mu
\end{pmatrix}
,
\end{split}
\ee
where the radial functions have the following boundary conditions:
\be
\begin{split}
\lim_{r\rightarrow 0}\frac{f\left(r\right)}{r}=0,\ \ \ \ \ \ \ \ \ \ \ \ \ \ \ \ \ \ \ \ \ \ \ \ \ \ \ \ \ \ \lim_{r\rightarrow\infty}f\left(r\right)=1,\\
\lim_{r\rightarrow 0}h\left(r\right)=0,\ \ \ \ \ \ \ \ \ \ \ \ \ \ \ \ \ \ \ \ \ \ \ \ \ \ \ \ \ \ \lim_{r\rightarrow\infty}h\left(r\right)=1.
\end{split}
\ee 
The field $B $ is an                                                                $SU(2)$-valued one-form,
\be
B\left(\mu,r,\theta,\phi\right)=B_rdr+B_{\theta}d\theta+B_{\phi}d\phi= B_idx^i,
\ee
for which we impose the radial gauge condition $B_r=0$ \cite{manton83}. We assume that in the radial gauge there exists a limiting field
\be\label{higgsinfty}
\Phi^{\infty}\left(\theta,\phi\right)\equiv \lim_{r\rightarrow\infty}\Phi\left(r,\theta,\phi\right)\;\;,
\ee
such that $\left|\Phi^{\infty} \right|=1$ and
\be
\Phi^{\infty}\left(\theta=0\right)=
\begin{pmatrix}
0\\
1\\
\end{pmatrix}.
\ee

\if
When $\theta_w\neq0$, the $U(1)$ field destroys spherical symmetry. However, the new equations retain axial symmetry about the z-axis. To this end, consider the set of orthonormal vectors \cite{manton78,rebbi80,kunz94}

\be
\begin{split}
\hat{e}_1(\phi)&=\left(\cos\phi,\sin\phi,0\right)\\ \hat{e}_2(\phi)&=\left(0,0,1\right)\\ \hat{e}_3(\phi)&=\left(\sin\phi,-\cos\phi,0\right)\;\;,\\
\end{split}
\ee
and the $SU(2)$-valued matrices
\be
M_m(\phi)= e_m ^a(\phi) \tau^a\;\;.
\ee
Using these, one can write the Ansatz for the $SU(2)_L \times U(1)_Y$ sphaleron. This wil be similar to the Ansatz given by \cite{manton78} for multimonopole solutions of $SU(2)$ gauge theory. The sphaleron Ansatz can be written as \cite{kunz91,kunz94}
\be\label{Axial Ansatz}
\begin{split}
B^i(\vec{r})&=B^i _a(\vec{r})\tau^a = u_m ^i (\phi)M_n(\phi)w_m ^n (\rho,z)\\
B^0(\vec{r})&= B^0 _a\tau ^a =0\\
A^i(\vec{r})&= u_m ^i(\phi) a_m(\rho,z)\\
A^0(\vec{r})&=0\\
\Phi(\vec{r})&=\frac{v}{\sqrt{2}}\left[h_0(\rho,z)+ih_m(\rho,z)M_m(\phi)\right]
\begin{pmatrix}
0\\
1\\ 
\end{pmatrix}\;\;. \\ 
\end{split}
\ee
The above Ansatz has axial symmetry, is parity invariant and is made up of 16 arbitrary real functions of $\rho$ and $z$ \cite{kunz94}. A suitable parameterization of the arbitrary functions that allows for the Ansatz to smoothly transform to the $SU(2)_L$ sphaleron Ansatz \cite{manton83,klink84} in the limit of vanishing mixing angle can be found in \cite{kunz92,kunz94}.
\fi
\if
section{Chern-Simons Charge and Baryon Number}
Consider the baryon current for a single generation of quarks,
\be
j^{\mu}_B = \frac{1}{3}\left( \bar{u}_ {\alpha}\gamma^{\mu}u_{\alpha} + \bar{d}_ {\alpha}\gamma^{\mu}d_{\alpha} \right),
\ee
where $\alpha$ is the $SU(3)$ color index. As a result of the Abelian anomaly in the standard model \cite{thooft76}, the non-vanishing divergence of this current has an SU(2) contribution given by \cite{klink84}
\be\label{baryon}
\partial_{\mu}j^{\mu}_B = \frac{g^2}{64\pi^2}\epsilon_{\mu\nu\rho\sigma} G^{\mu\nu}_aG^{\rho\sigma}_a.
\ee 
Furthermore, the right-hand side of Eq.(\ref{baryon}) can also be written as the total divergence of the gauge variant Chern-Simons current given by 
\be
K_{\mu}=\frac{g^2}{16\pi^2}\epsilon_{\mu\nu\rho\sigma}\text{Tr}\left(G^{\nu\rho}B^{\sigma} + \frac{2}{3}igB^{\nu}B^{\rho}B^{\sigma}\right),
\ee
where
\be
G_{\nu\rho}=\frac{1}{2}\tau^a G^a _{\nu\rho},\;\;\;B_{\nu}=\frac{1}{2}\tau^a B^a _{\nu}.
\ee
The Chern-Simons charge of a field configuration is defined as
\be\label{Q_CS}
Q_{\text{CS}}=\int d^3r K^0.
\ee
When calculated in the correct gauge, namely one in which the integral of $\vec{K}.\vec{dS}$ over the surface of a sphere $S$ at spatial infinity vanishes, Eqs.(\ref{baryon}-\ref{Q_CS}) imply that the Chern-Simons charge of the field configuration is equal to its baryon (lepton) number $Q_B$ ($Q_L$). For the $SU(2)_L$ sphaleron of YMH theory, if one uses a one-point compactification for 3-space and starts from a vacuum configuration with $Q_{CS}$ set to zero and traverses the NCL through the sphaleron, one finds that the sphaleron will have $Q_B= Q_{CS}=\frac{1}{2}$ \cite{klink84}, while the neighboring vacuum will have $Q_{CS}=1$. Furthermore, even when $\theta_w\neq0$, since the electric and magnetic fields are perpendicular at the sphaleron, the $U(1)$ field does not contribute to the baryon number. Therefore, for the axially symmetric Ansatz \cite{kunz92} of the $SU(2)_L \times U(1)_Y$ sphaleron, once again $Q_B=\frac{1}{2}$. Since the above argument also applies to leptonic currents, the electroweak sphaleron's lepton number is also $\frac{1}{2}$. 
In the next section, we augment CP and present an alternative derivation of the fermion numbers of the S sphaleron based on discrete symmetries reflected in the spectral mirror symmetry of the Dirac Hamiltonian.
\fi
\section{CP Transformation along NCL}\label{CP Transformation along Noncontractible Loop}

In this section we study the behavior of the Dirac Hamiltonian under discrete transformations including C and P in a sphaleron background. 
When the weak mixing angle goes to zero, we perform our analysis for arbitrary loop parameter $\mu$. For the $SU(2)_L\times U(1)_Y$ sphaleron, only the sphaleron Ansatz has been constructed and not the full barrier. This restricts our analysis to the sphaleron when $\theta_w=0$. Nevertheless, this strategy can be readily extended to the full barrier once it is constructed. 
\if
Let us first briefly consider a 1+1 dimensional theory of effectively massive fermions interacting nonlinearly with a pseudoscalar field. The Lagrangian for this theory is given by \cite{mackenzie84}
\be 
\mathcal{L}=\bar{\psi}\left(i\gamma^{\mu} \partial_{\mu} - me^{i\phi(x)\gamma^5}\right)\psi,
\ee
with the topologically nontrivial background field given by
\be\label{soliton}
\phi(x)=\mu \;\frac{x}{\left| x\right|},\;\;\;\;\;\;\;\;\mu\ \in \left(0,\pi\right).
\ee
This model, which was studied in detail by MacKenzie and Wilczek in \cite{mackenzie84}, is in fact a chirally rotated, infinitely thin version of the one studied in \cite{jackiw76}. Here, we are interested in the behavior of the Dirac Hamiltonian operator under fermion number conjugation. The Hamiltonian operator is
\be 
\hmch = -i\gamma^0\gamma^j\partial_j + m\gamma^0 e^{i\phi(x)\gamma^5}
\ee
and the choices of gamma matrices and the resulting charge conjugation operator are \cite{mackenzie84}
\be
\begin{split}
\gamma^0 = \sigma^1 ,\;\;\;\;\; \gamma^1 = i\sigma^3 ,\;\;\;\;\;  \gamma^5 = \gamma^0 \gamma^1 = \sigma^2 ,\;\;\;\;\; \psi^C(x) = \sigma^1 \psi^*(x).
\end{split}
\ee
Inserting the expression for $\phi(x)$ given by Eq.(\ref{soliton}), the charge-conjugated Hamiltonian operator in this representation becomes
\be
\hmch^C \equiv C^{-1} \hmch C = i\gamma^0\gamma^j\partial_j + m\left(\cos 2\mu\; \gamma^0 - i\frac{\left | x \right |}{x} \sin 2\mu \;\gamma^1 \right) e^{i\phi(x)\gamma^5}.
\ee
This implies that precisely at $\mu = \frac{\pi}{2}$ the Dirac Hamiltonian is odd under charge conjugation, i.e.
\be
\hmch^C\left(x,t,\mu=\frac{\pi}{2}\right) = - \hmch\left(x,t,\mu=\frac{\pi}{2}\right).
\ee
This in turn implies that for every eigenstate with energy $E$ there is one with energy $-E$, and hence the spectrum has mirror symmetry.
An important observation to make in the analysis of \cite{mackenzie84} is that, as the nontrivial classical configuration adiabatically forms from the trivial one, a bound state separates from the positive continuum at $\mu=0$, crosses $E=0$ at $\mu = \frac{\pi}{2}$ and joins the Dirac sea at $\mu=\pi$. Meanwhile, the spectral deficiency $\mathcal{D}$ in the positive continuum caused by the bound state starts to replenish, while deficiency starts to build up in the Dirac sea. At $\mu = \frac{\pi}{2}$ the fermionic bound state is at $E=0$, and the spectral deficiencies in both continua are \cite{gousheh94}
\be
\mathcal{D} = \frac{\mu}{\pi}.
\ee
Therefore, at $\mu = \frac{\pi}{2}$, the quantum field theoretic expectation value of the fermion number operator is \cite{mackenzie84}
\be
\left| \left\langle N \right\rangle \right| = \frac{1}{2}.
\ee
This number can be interpreted as the fermion number of the bosonic configuration.
\fi

Consider the Dirac Hamiltonian operator of the standard electroweak theory at $\theta_w=0$ \cite{klink03}
\be\label{loop hamiltonian}
\hmch= -i\gamma^0 \gamma^j D_j + k\gamma^0\left(\Phi_M^{\dagger}P_L + \Phi_M P_R\right),
\ee
where the matrix $\Phi_M$ contains the scalar fields of the Higgs doublet and its charge-conjugated doublet and is given by
\be\label{Phi_M}
\Phi_M = 
\begin{pmatrix}
\begin{array}{cc}
\phi_2^* & \phi_1\\
-\phi_1^*& \phi_2\\
\end{array}
\end{pmatrix},
\ee
and the projection operators are defined as
\be
P_L=\frac{1}{2}\left(1-\gamma^5\right),\;\;\;\;\;P_R=\frac{1}{2}\left(1+\gamma^5\right).\ee
We now use the Ansatz given in Eq.(\ref{hedgehog Ansatz}) to construct $\Phi_M$ shown in Eq.(\ref{Phi_M}) and insert it into Eq.(\ref{loop hamiltonian}) to obtain the expression for $\hmch$ along the NCL. The final expression for $\hmch$ is shown in the appendix.

We use the following choice of Weyl representation for our gamma matrices
\be
\gamma^0 =
\begin{pmatrix}
\begin{array}{cc}
0 & I_2 \\
I_2 & 0 \\ 
\end{array}
\end{pmatrix}
,\;\;\;
\gamma^i =
\begin{pmatrix}
\begin{array}{cc}
0 & \sigma^i \\
-\sigma^i & 0\\
\end{array}
\end{pmatrix}
,\;\;\;
\gamma^5 = 
\begin{pmatrix}
\begin{array}{cc}
-I_2 & 0 \\
0 & I_2 \\
\end{array}
\end{pmatrix}.
\ee
In this representation, charge conjugation acts non-trivially on scalars and spinors, both of which transform in the fundamental representation of $SU(2)$, as
\be
\Phi\left(\vec{x},t\right) \xrightarrow{C} i\tau^2 \Phi^*\left(\vec{x},t\right) ,\;\;\;\;\;\;\;\;\;\;\;\;\; \Psi\left(\vec{x},t\right) \xrightarrow{C} i\tau^2 \gamma^2 \Psi^*\left(\vec{x},t\right),
\ee
while under the combined transformation of C and P,
\be
\Phi\left(\vec{x},t\right) \xrightarrow{CP} i\tau^2 \Phi^*\left(-\vec{x},t\right) ,\;\;\;\;\;\;\;\;\;\;\;\;\; \Psi\left(\vec{x},t\right) \xrightarrow{CP} i\tau^2 \gamma^2 \gamma^0 \Psi^*\left(-\vec{x},t\right).
\ee
Therefore, the charge-conjugated, parity-inverted Hamiltonian becomes
\be\label{hcp}
\hmch^{CP}\left(\vec{x},t,\mu\right) = \gamma^2\gamma^0 
\begin{pmatrix}
\begin{array}{cc}
-\hmch_{22}^* & \hmch_{21}^*\\
\hmch_{12}^* & -\hmch_{11}^*\\
\end{array}
\end{pmatrix}_{(-\vec{x},t,\mu)}
\gamma^0\gamma^2.
\ee

After inserting the explicit expressions for the matrix elements of $\hmch$ given in Eq.(\ref{components}) into Eq.(\ref{hcp}), we observe that nowhere along the NCL is $\hmch$ odd under CP, except at the trivial vacuum ($\mu=0,\pi$). However, at $\mu=\frac{\pi}{2}$, there are many cancellations and the even part reduces to
\be\label{hamiltonian sum}
\begin{split}
\hmch^{CP}&\left(\vec{x},t,\mu=\frac{\pi}{2}\right) + \;\hmch\left(\vec{x},t,\mu=\frac{\pi}{2}\right) \\
&= 2k\eta h(r)\gamma^0
\begin{pmatrix}
\begin{array}{cc}
\cos\theta \left(P_L + P_R \right) & 
-\sin\theta e^{i\phi} \left(P_L - P_R \right)\\
\sin \theta e^{-i\phi} \left(P_L - P_R \right) &
\cos \theta \left(P_L + P_R \right) \\ 
\end{array}
\end{pmatrix}.
\end{split}
\ee
This reflects the fact that the pseudoscalar sphaleron configuration breaks the CP invariance of the one-generation electroweak theory that we have been considering. We now define a new conjugation transformation, $\widetilde{CP}$, which consists of CP and is augmented by an additional operation as follows
\be\label{CP tilde}
\widetilde{CP} \equiv CP \mathcal{X},
\ee 
where $\mathcal{X}= e^{-i\mu\gamma^5}$. By repeating the calculation leading to Eq.(\ref{hamiltonian sum}) for the new operation, Eq.(\ref{CP tilde}), we see that
\be\label{hamiltonian sum new}
\hmch^{\widetilde{CP}}\left(\vec{x},t,\mu=\frac{\pi}{2}\right) = - \;\hmch\left(\vec{x},t,\mu=\frac{\pi}{2}\right).
\ee

The existence of a transformation under which $\hmch$ is odd ascertains the spectral mirror symmetry. That is, under such a transformation every eigenstate of $\hmch$ with energy $E$ is transformed into one with energy $-E$, the only exception being a zero energy bound state which must then be invariant under such a transformation. From the viewpoint of spectral deficiency \cite{gousheh94}, this implies that as the NCL is traversed, spectral deficiency $\mathcal{D}$ in the positive continuum caused by the bound state starts to replenish, while deficiency starts to build up in the Dirac sea. At $\mu = \frac{\pi}{2}$ the fermionic bound state is at $E=0$, and the spectral deficiencies in both continua are \cite{gousheh94}
\be
\mathcal{D} = \frac{\mu}{\pi}.
\ee
Therefore, at $\mu = \frac{\pi}{2}$, the quantum field theoretic expectation value of the fermion number operator is \cite{mackenzie84}
\be
\left| \left\langle N \right\rangle \right| = \frac{1}{2}.
\ee
This number can be interpreted as the fermion number of the bosonic configuration. 

In the next section, we check the invariance of the zero mode presented by Ringwald (which is the one that is relevant to our model) \cite{ringwald88} under $\widetilde{CP}$. The fermion numbers $\frac{1}{2}$ of the sphaleron then follow immediately from the reasoning presented by Jackiw and Rebbi or, equivalently, from the spectral deficiency $\frac{1}{2}$ of the Dirac sea. It is worth mentioning that at the trivial vacuum ($\mu=0,\pi$), $\hmch$ is odd under CP, showing that the spectrum has mirror symmetry there. However, there are no bound states in the trivial vacuum.
\if
Let us now concentrate on the chiral anomaly which leads to fermion number violation, given in Eq.(\ref{baryon}), in the symmetric phase. The chiral current $j^{\mu}_5 =\bar{\Psi}\gamma^{\mu}\gamma^5\Psi$ for a single isodoublet $\Psi$ of left-handed fermions has the following anomaly 
\be\label{chiralcurrent}
\partial_{\mu}j^{\mu}_5 = \frac{g^2}{32\pi^2}\epsilon_{\mu\nu\rho\sigma} G^{\mu\nu}_aG^{\rho\sigma}_a.
\ee
Upon integrating both sides of Eq.(\ref{chiralcurrent}) over spacetime along the NCL, with the loop parameter $\mu$ playing the role of (imaginary) time, we arrive at the Atiyah-Singer index theorem \cite{atiyah},
\be\label{intchiral}
N_{+} - N_{-} = -\int d^4x \mathcal{P}(x),
\ee
where $N_+$ and $N_-$ are the number of chiral states that cross $E=0$ from above and below, respectively. In the level-crossing picture for a single left-handed fermion doublet coupled to the $SU(2)$ fields of the NCL, as the path from one vacuum to a neighboring vacuum is traversed, one unit of left-handed charge arises from the negative energy continuum and reaches the positive energy continuum at the next vacuum. At the sphaleron, this fermionic state is a zero mode. This has been interpreted as a violation of fermionic charge by one unit \cite{klink03}. Since baryons and leptons have the same behavior along the NCL with respect to the chiral anomaly, this implies that the sphaleron process conserves $B-L$ but violates $B+L$, where $B$ and $L$ are the baryon and lepton numbers, respectively. 

Note that the Pontryagin density $\mathcal{P}(x)$ is odd under CP. Consider the NCL in the one-point compactified scheme. Regardless of their Chern-Simons numbers, the degenerate vacua, characterized by time-independent pure gauge configurations $gB_i = -\partial_i U U^{-1}$, are independently C, P and T invariant. Furthermore, as was mentioned in Section \ref{SU2Sphaleron}, excluding the sphaleron, the rest of the configurations along the NCL have no definite parity. This leaves us with the sphaleron sitting at the top of the NCL. Here, as was shown in Eq.(\ref{hamiltonian sum new}), we see that it is the combination of chirality and CP transformation that leads to spectral mirror symmetry.
\fi
\if
}
\fi
\if
At this point, one should remember that in the analysis of \cite{jackiw76}, emphasis was placed on the fermion number conjugation invariance of the zero mode and not in general on the entire spectrum. This motivates us to take a closer look at the problem. In this respect, consider the eigenvalue problem for the Hamiltonian given by Eq. \ref{loop hamiltonian},
\be
\hmch \Psi = E \Psi.
\ee
Doing so for field configurations along the sphaleron barrier leads us to the following useful theorem:

\bigskip
\noindent \textbf{Theorem:} \textit{For the gauge and Higgs fields of the NCL, a $\widetilde{CP}$-invariant fermionic zero mode is possible only at the sphaleron.}

\bigskip
\noindent \textbf{Proof:} Suppose $E$ is an eigenvalue of Eq. \ref{loop hamiltonian} whose components are given by Eq. \ref{components}. Recall the transformation properties of spinor doublets under C and P. For the first component $\psi_1$ of the spinor eigendoublet we can write
\begin{align*}
E &\left(\psi_1 + \psi^{\widetilde{CP}}_1 \right) = \\ 
& \left[-if(r)r \gamma^0 \gamma^1 P_L e^{i(\mu+\phi)} \sin^2\mu\cos\theta \left( \sin^2\theta\sin\phi + i\cos\phi\right) \right.\\ 
& +if(r)r \gamma^0 \gamma^2 P_L e^{i(\mu +\phi)} \sin^2 \mu \cos\theta \left( \sin^2 \theta \cos\phi - i \sin\phi \right) \\
& -f(r)r\gamma^0 \gamma^3 P_L e^{i(\mu+\phi)}\sin^2\mu\sin\theta  \\
& \left.-k\eta h(r)\gamma^0\sin\mu\sin\theta e^{i\phi}\left(P_L-P_R\right)\vphantom{-if(r)} \right] \left(\psi_2 - \psi^{\widetilde{CP}}_2 \right)\\
& +\cos\mu\left[if(r)r\gamma^0\gamma^1P_L e^{i(\mu +\phi)} \sin\mu \left(\cos^2\theta\cos\phi - i\sin^2\theta\sin\phi \right) \right. \\
& +if(r)r\gamma^0\gamma^2P_L e^{i(\mu +\phi)} \sin\mu \left(\cos^2\theta\sin\phi + i\sin^2\theta\cos\phi \right) \\
& \left.-if(r)r\gamma^0\gamma^3P_L e^{i(\mu +\phi)} \sin\mu\sin\theta\cos\theta \vphantom{+if(r)}\right] \left(\psi_2 + \psi^{\widetilde{CP}}_2\right) \\
& +\left[-i\gamma^0\gamma^j\partial_j + f(r)r\gamma^0\gamma^1P_L\sin^2\mu\sin^3\theta\sin\phi \right. \\
& \left. -f(r)r\gamma^0\gamma^2P_L\sin^2\mu\sin^3\theta\cos\phi + ik\eta h(r)\gamma^0\sin\mu\cos\theta\left(e^{-i\mu}P_L - e^{i\mu}P_R\right)\vphantom{-f(r)}\right]\left(\psi_1 - \psi^{\widetilde{CP}}_1 \right) \\
& +\cos\mu\left[-f(r)r\gamma^0\gamma^1P_L\sin\mu\sin\theta\cos\theta\cos\phi \right. \\
& -f(r)r\gamma^0\gamma^2P_L\sin\mu\sin\theta\cos\theta\sin\phi + f(r)r\gamma^0\gamma^3P_L\sin\mu\sin^2\theta \\
& +\left.k\eta \gamma^0\left(e^{-i\mu}P_L + e^{i\mu}P_R\right)\vphantom{-f(r)}\right]\left(\psi_1 + \psi^{\widetilde{CP}}_1 \right). \numberthis \label{energy sum} \\
\end{align*}
Let $E=0$. Then the left-hand side of the equation is $0$. Now let $\Psi^{\widetilde{CP}} = \Psi$; then the first and third terms of the right-hand side also vanish. This implies that $\cos\mu=0$. Thus $\mu = \frac{\pi}{2}$. A similar argument for the second component $\psi_2$ of the spinor eigendoublet also holds, which completes the proof. $\square$
\fi
\section{The Zero Mode}
\label{The Zero Mode}

Recall that in the original analysis of Jackiw and Rebbi, the fermionic zero mode in the soliton background was fermion number self-conjugate \cite{jackiw76}. Thus, an important consistency check on our symmetry transformation would be to operate it on the fermionic zero mode that was given by Ringwald at the electroweak S sphaleron \cite{ringwald88}. To this end, consider the zero-energy solution of the Dirac equation in the sphaleron background. The Ansatz for the left-handed isodoublet of the zero mode is given by \cite{ringwald88}
\be\label{zero mode psi}
\Psi_{0,L}^{i\alpha}\left(\vec{x},t\right) = \epsilon^{i\alpha} z(r),
\ee
where $i=1,2$ is the weak isospin index, $\alpha=1,2$ is the spinor index and $\epsilon^{ij}$ is the Levi-Civita symbol ($\epsilon^{12}= +1$). The functional form of $z(r)$ is obtained by solving the radial part of the Dirac equation. Depending on whether the fermions are massive or massless, $z(r)$ will take on a different form \cite{ringwald88}. For a single generation of left-handed quarks, denoted by
\be
\Psi_{0,L}^{\alpha} = 
\begin{pmatrix}
u_{0,L}^{\alpha} \\
d_{0,L}^{\alpha}\\
\end{pmatrix},
\ee
this implies that
\be
u_{0,L}\left(\vec{x},t\right) = z(r)\begin{pmatrix}
0\\1
\end{pmatrix}
\equiv z(r)\ket{\downarrow},\;\;\;\;\;
d_{0,L}\left(\vec{x},t\right) = z(r) \begin{pmatrix}
-1\\0\\
\end{pmatrix}
\equiv -z(r)\ket{\uparrow}.
\ee
Thus, Eq.(\ref{zero mode psi}) can also be written as
\be\label{spin zero mode}
\Psi_{0,L}\left(\vec{x},t\right) = z(r)\begin{pmatrix}
\ket{\downarrow} \\
-\ket{\uparrow} \\
\end{pmatrix}.
\ee
A CP transformation on Eq.(\ref{zero mode psi}) or, equivalently, Eq.(\ref{spin zero mode}) yields
\be
\Psi_{0,L}^{CP}\left(\vec{x},t\right) = i\gamma^5 \Psi_{0,L}\left(\vec{x},t\right),
\ee
which shows that, as expected, the zero mode is not CP-invariant. By noting that we are performing the symmetry transformation at the sphaleron ($\mu = \frac{\pi}{2}$), implementing the additional factor of $-i\gamma^5$ required by a $\widetilde{CP}$ transformation, we obtain 
\be
\Psi_{0,L}^{\widetilde{CP}}\left(\vec{x},t\right) = \Psi_{0,L}\left(\vec{x},t\right).
\ee
Thus, we observe that the fermionic zero mode of Ringwald in the sphaleron background is $\widetilde{CP}$-invariant.
\if In fact, this is the only configuration along the NCL where this is possible. To see this consider the relation
\be\label{theorem}
\hmch \left(\Psi_{n,L} + \Psi_{n,L}^{\widetilde{CP}} \right) = E_n \left(\Psi_{n,L} - \Psi_{n,L}^{\widetilde{CP}} \right),
\ee
where $E_n$ is the energy eigenvalue of the state $\Psi_{n,L}$. Inserting Eq.(\ref{components}) into Eq.(\ref{theorem}), it is not difficult to verify that demanding a $\widetilde{CP}$ invariant zero mode singles out the sphaleron at $\mu=\frac{\pi}{2}$. 
\fi
\if
We now turn to the more physical case of non-vanishing mixing angle. There are two points to consider for such an analysis: first, the full sphaleron barrier for the case of a non-zero mixing angle is yet to be parametrically  constructed; this means that we can only perform our analysis at the sphaleron itself. Second, in an axially symmetric Ansatz, the arbitrary functions in \ref{Axial Ansatz} will in general not be parity invariant. This is in contrast with the hedgehog Ansatz, where the functions only had radial dependence. What we can do in this case is derive constraints on the arbitrary functions  of the Ansatz.

Since the number of fermion generations and fermion type (quark or lepton) do not affect our symmetry analysis, for the sake of notational ease, let us consider only the first generation of quarks. As far as the standard electroweak theory is concerned, however, the argument can be readily extended to include the full fermionic sector of the theory. In the presence of the Abelian field, the Dirac Hamiltonian for static gauge fields reads
\be\label{axial Hamiltonian}
\begin{split}
\mch =& -\bar{\psi}_L i\gamma^j D_j^L - \bar{u}_R i\gamma^jD_j^Ru_R - \bar{d}_R i\gamma^j D_j^R d_R \\
&+ k\left [\bar{\psi}_L \left(i\Phi^C u_R + \Phi d_R \right) + \left( \bar{d}_R \Phi^{\dagger} - \bar{u}_R i(\Phi^C)^{\dagger}\right)\psi_L \right]\;,
\end{split}
\ee
where
\be
\begin{split}
D_j^L &=\; \partial_j - ig\frac{\tau^a}{2}B_j^a - i\frac{g^{\prime}}{2}Y^LA_j \\
D_j^R &=\; \partial_j - i\frac{g^{\prime}}{2}Y^RA_j \\
\Phi^C &= \tau^2\Phi^*\;\;,
\end{split}
\ee
$u_R$ and $d_R$ are $SU(2)$ singlets and $\psi_L$ is the $SU(2)$ doublet
\be
\psi_L =
\begin{pmatrix}
u_L \\
d_L \\
\end{pmatrix}\;\;.
\ee
Demanding CP invariance of the Dirac Hamiltonian only at the sphaleron configuration of a possible NCL imposes constraints on the arbitrary functions of the Ans\ddot{a}tze at the sphaleron. To see this, consider the term
\be 
\hmch^R \equiv -i\gamma^0 \gamma^jD_j ^R = -i\gamma^0 \gamma^j \left(\partial_j - i \frac{g^{\prime}}{2} Y A_j \right) \;\;.
\ee
Inserting \ref{Axial Ansatz} into the above equation and noting that under parity
\be
\vec{u}_1 \rightarrow -\vec{u}_1,\;\;\;\;\;\;\;\;\;\;\vec{u}_2 \rightarrow \vec{u}_2,\;\;\;\;\;\;\;\;\;\; \vec{u}_3 \rightarrow -\vec{u}_3,
\ee
we find that demanding CP invariance at the sphaleron requires the functions $a^i(\rho, z)$ to transform under parity as
\be
a^1 \rightarrow a^1,\;\;\;\;\;\;\;\;\;\;\;
a^2 \rightarrow -a^2,\;\;\;\;\;\;\;\;\;\;
a^3 \rightarrow a^3.  
\ee
Imposing fermion number conjugation on the other terms of the Dirac Hamiltonian operator will impose similar constraints on the functions $w^n _m$ and $h_m$ as well.
\fi
\section{Summary and Discussion}\label{Summary and Discussion}
In this paper, we have studied the behavior of fermions under discrete transformations in a sphaleron background. 
For the fields of the NCL passing through the S sphaleron, it is well known that the system is not CP-invariant except at the vacua. However, we have constructed a new transformation, denoted by $\widetilde{CP}$, by augmenting a CP transformation with an additional operation that acts nontrivially in the Yukawa sector and has the following important property. We see that for field configurations along the NCL, the Dirac Hamiltonian is odd under $\widetilde{CP}$ precisely at the S sphaleron sitting at the top of the barrier that begins and ends at the trivial vacuum. This ensures that the spectrum has mirror symmetry. That is, for every positive energy eigenstate there is a corresponding negative energy one and the zero mode, if any, is self-conjugate.

As an important consistency check, by performing the symmetry transformation on the fermionic zero mode given by Ringwald \cite{ringwald88} in the sphaleron background, we observe that the zero mode is indeed $\widetilde{CP}$-invariant. This is closely analogous to the analyses of \cite{jackiw76, mackenzie84}. There, fermion number conjugation symmetry of the spectrum including the zero mode in the background of the classical solution was an important condition that led to the derivation of the half-integer fermion numbers of the background bosonic fields. In the analyses of \cite{jackiw76, mackenzie84}, fermion number conjugation was charge conjugation. Our transformation operator is $\widetilde{CP}$ which reveals the spectral mirror symmetry at the sphaleron. In this configuration, the spectral deficiency in the Dirac sea is exactly $\frac{1}{2}$ and one associates this to the fermion number of the background field which is the sphaleron.

Overall, this analysis offers a number of other potential advantages. At a basic
level, it can provide a useful consistency check for the numerous fractionally-charged sphaleron Ans\"{a}tze that have been discovered so far \cite{kunz08, kunz08b, kunz04, kleihaus99, kunzmulti94, klink93, klink05, klink17}, and helps place constraints on their functional forms. An example of this can be seen in the Ansatz for the axially symmetric sphaleron, where the arbitrary functions acquire a z-dependence \cite{kunz91}. Furthermore, the analyses of \cite{jackiw76, mackenzie84} required C-invariance, while the present analysis led to $\widetilde{CP}$-invariance. It may be that other solutions require other novel symmetry transformations for the fermionic sector to correctly explain their fractional charges. An important issue that our study has not addressed is what happens when one considers three generations of fermions, where CP symmetry is violated through the CKM and PMNS mixing matrices in the background of the even-parity Higgs field vacuum. 

Finally, from a more practical perspective, one should bear in mind that sphalerons currently play an omnipresent role in physics and show up in many field theories, such as gravitation, electroweak theory and quantum chromodynamics. Thus, it seems worthy to delve even deeper into their structure to see if new symmetries emerge. It may be that studying these symmetries paves the way for a more systematic understanding of the topological properties of unstable solutions in gauge field theories and their physical applications. 

\bigskip
\noindent Acknowledgments: M. M. would like to thank N. Manton and J. Kunz for useful discussions and N. Dadkhah for useful comments on the manuscript. We would like to thank the research council of the Shahid Beheshti University for financial support.

\appendix

\section{Dirac Hamiltonian along NCL}

In this section we give the explicit functional form of the components of Eq.(\ref{loop hamiltonian}). As an $SU(2)$-valued $2\times 2$ matrix, $\hmch$ is
\be
\hmch = 
\begin{pmatrix}
\begin{array}{cc}
\hmch_{11}& \hmch_{12}\\
\hmch_{21}& \hmch_{22}\\
\end{array}
\end{pmatrix}.
\ee
In the background of the gauge and Higgs fields of the NCL, Eq.(\ref{hedgehog Ansatz}), the components of $\hmch$ are
\begin{subequations}\label{components}
\begin{align}
\begin{split}
\hmch_{11} = &-i\gamma^0\gamma^j\partial_j + f(r)r\gamma^0\gamma^3P_L\sin\mu\cos\mu\sin^2\theta \\  &- f(r)r\gamma^0\gamma^1P_L\sin\mu\sin\theta\left(\cos\mu\cos\theta\cos\phi - \sin\mu\sin^2\theta\sin\phi\right) \\ &- f(r)r\gamma^0\gamma^2P_L\sin\mu\sin\theta\left(\cos\mu\cos\theta\sin\phi + \sin\mu\sin^2\theta\cos\phi\right) \\& + k\eta h(r)\gamma^0\left[e^{-i\mu} \left( \frac{\cos\mu}{h(r)} + i\sin\mu\cos\theta\right)P_L + e^{i\mu}\left(\frac{\cos\mu}{h(r)} - i\sin\mu\cos\theta\right) P_R \right],
\end{split}
\\[2ex]
\begin{split}
\hmch_{12} = &\;if(r)r\gamma^0\gamma^1P_Le^{i(\mu+\phi)}\sin\mu\\
&\times\left[\cos\theta\cos\phi\left(\cos\mu\cos\theta-i\sin\mu\right)-i\sin^2\theta\sin\phi\left(\cos\mu-i\sin\mu\cos\theta\right)\right]\\ +&\; if(r)r\gamma^0\gamma^2P_Le^{i(\mu+\phi)}\sin\mu\\
&\times\left[\cos\theta\sin\phi\left(\cos\mu\cos\theta-i\sin\mu\right)+i\sin^2\theta\cos\phi\left(\cos\mu-i\sin\mu\cos\theta\right)\right]\\
-&\;if(r)r\gamma^0\gamma^3P_Le^{i(\mu+\phi)}\sin\mu\sin\theta\left(\cos\mu\cos\theta-i\sin\mu\right)\\
-&\;k\eta h(r)\gamma^0\sin\mu\sin\theta e^{i\phi}\left(P_L - P_R\right),
\end{split} 
\end{align}
\begin{align}
\begin{split}
\hmch_{21} = &\;-if(r)r\gamma^0\gamma^1P_Le^{-i(\mu+\phi)}\sin\mu\\
&\times\left[\cos\theta\cos\phi\left(\cos\mu\cos\theta+i\sin\mu\right)+i\sin^2\theta\sin\phi\left(\cos\mu+i\sin\mu\cos\theta\right)\right]\\ -&\; if(r)r\gamma^0\gamma^2P_Le^{-i(\mu+\phi)}\sin\mu\\
&\times\left[\cos\theta\sin\phi\left(\cos\mu\cos\theta+i\sin\mu\right)-i\sin^2\theta\cos\phi\left(\cos\mu+i\sin\mu\cos\theta\right)\right]\\
+&\;if(r)r\gamma^0\gamma^3P_Le^{-i(\mu+\phi)}\sin\mu\sin\theta\left(\cos\mu\cos\theta+i\sin\mu\right)\\
+&\;k\eta h(r)\gamma^0\sin\mu\sin\theta e^{-i\phi}\left(P_L - P_R\right),
\end{split}
\\[2ex]
\begin{split}
\hmch_{22} = &-i\gamma^0\gamma^j\partial_j - f(r)r\gamma^0\gamma^3P_L\sin\mu\cos\mu\sin^2\theta \\  &+ f(r)r\gamma^0\gamma^1P_L\sin\mu\sin\theta\left(\cos\mu\cos\theta\cos\phi - \sin\mu\sin^2\theta\sin\phi\right) \\ &+ f(r)r\gamma^0\gamma^2P_L\sin\mu\sin\theta\left(\cos\mu\cos\theta\sin\phi + \sin\mu\sin^2\theta\cos\phi\right) \\& + k\eta h(r)\gamma^0\left[e^{+i\mu} \left( \frac{\cos\mu}{h(r)} - i\sin\mu\cos\theta\right)P_L + e^{-i\mu}\left(\frac{\cos\mu}{h(r)} + i\sin\mu\cos\theta\right) P_R \right].
\end{split} 
\end{align}
\end{subequations}

\nocite{apsrev41Control}
\bibliographystyle{apsrev4-1}
\bibliography{paper1bib}

\begin{thebibliography}{44}%
\makeatletter
\providecommand \@ifxundefined [1]{%
 \@ifx{#1\undefined}
}%
\providecommand \@ifnum [1]{%
 \ifnum #1\expandafter \@firstoftwo
 \else \expandafter \@secondoftwo
 \fi
}%
\providecommand \@ifx [1]{%
 \ifx #1\expandafter \@firstoftwo
 \else \expandafter \@secondoftwo
 \fi
}%
\providecommand \natexlab [1]{#1}%
\providecommand \enquote  [1]{``#1''}%
\providecommand \bibnamefont  [1]{#1}%
\providecommand \bibfnamefont [1]{#1}%
\providecommand \citenamefont [1]{#1}%
\providecommand \href@noop [0]{\@secondoftwo}%
\providecommand \href [0]{\begingroup \@sanitize@url \@href}%
\providecommand \@href[1]{\@@startlink{#1}\@@href}%
\providecommand \@@href[1]{\endgroup#1\@@endlink}%
\providecommand \@sanitize@url [0]{\catcode `\\12\catcode `\$12\catcode
  `\&12\catcode `\#12\catcode `\^12\catcode `\_12\catcode `\%12\relax}%
\providecommand \@@startlink[1]{}%
\providecommand \@@endlink[0]{}%
\providecommand \url  [0]{\begingroup\@sanitize@url \@url }%
\providecommand \@url [1]{\endgroup\@href {#1}{\urlprefix }}%
\providecommand \urlprefix  [0]{URL }%
\providecommand \Eprint [0]{\href }%
\providecommand \doibase [0]{http://dx.doi.org/}%
\providecommand \selectlanguage [0]{\@gobble}%
\providecommand \bibinfo  [0]{\@secondoftwo}%
\providecommand \bibfield  [0]{\@secondoftwo}%
\providecommand \translation [1]{[#1]}%
\providecommand \BibitemOpen [0]{}%
\providecommand \bibitemStop [0]{}%
\providecommand \bibitemNoStop [0]{.\EOS\space}%
\providecommand \EOS [0]{\spacefactor3000\relax}%
\providecommand \BibitemShut  [1]{\csname bibitem#1\endcsname}%
\let\auto@bib@innerbib\@empty
\bibitem [{\citenamefont {Jackiw}\ and\ \citenamefont
  {Rebbi}(1976)}]{jackiw76}%
  \BibitemOpen
  \bibfield  {author} {\bibinfo {author} {\bibfnamefont {R.}~\bibnamefont
  {Jackiw}}\ and\ \bibinfo {author} {\bibfnamefont {C.}~\bibnamefont {Rebbi}},\
  }\bibfield  {title} {\enquote {\bibinfo {title} {Solitons with fermion number
  1/2},}\ }\href {\doibase 10.1103/PhysRevD.13.3398} {\bibfield  {journal}
  {\bibinfo  {journal} {Phys. Rev. D}\ }\textbf {\bibinfo {volume} {13}},\
  \bibinfo {pages} {3398--3409} (\bibinfo {year} {1976})}\BibitemShut {NoStop}%
\bibitem [{\citenamefont {Jackiw}(1977)}]{jackiw77}%
  \BibitemOpen
  \bibfield  {author} {\bibinfo {author} {\bibfnamefont {R.}~\bibnamefont
  {Jackiw}},\ }\bibfield  {title} {\enquote {\bibinfo {title} {Quantum meaning
  of classical field theory},}\ }\href {\doibase 10.1103/RevModPhys.49.681}
  {\bibfield  {journal} {\bibinfo  {journal} {Rev. Mod. Phys.}\ }\textbf
  {\bibinfo {volume} {49}},\ \bibinfo {pages} {681--706} (\bibinfo {year}
  {1977})}\BibitemShut {NoStop}%
\bibitem [{\citenamefont {Goldstone}\ and\ \citenamefont
  {Wilczek}(1981)}]{goldwilczek81}%
  \BibitemOpen
  \bibfield  {author} {\bibinfo {author} {\bibfnamefont {Jeffrey}\ \bibnamefont
  {Goldstone}}\ and\ \bibinfo {author} {\bibfnamefont {Frank}\ \bibnamefont
  {Wilczek}},\ }\bibfield  {title} {\enquote {\bibinfo {title} {Fractional
  quantum numbers on solitons},}\ }\href {\doibase 10.1103/PhysRevLett.47.986}
  {\bibfield  {journal} {\bibinfo  {journal} {Phys. Rev. Lett.}\ }\textbf
  {\bibinfo {volume} {47}},\ \bibinfo {pages} {986--989} (\bibinfo {year}
  {1981})}\BibitemShut {NoStop}%
\bibitem [{\citenamefont {MacKenzie}\ and\ \citenamefont
  {Wilczek}(1984{\natexlab{a}})}]{mackenzie84}%
  \BibitemOpen
  \bibfield  {author} {\bibinfo {author} {\bibfnamefont {R.}~\bibnamefont
  {MacKenzie}}\ and\ \bibinfo {author} {\bibfnamefont {Frank}\ \bibnamefont
  {Wilczek}},\ }\bibfield  {title} {\enquote {\bibinfo {title} {Illustrations
  of vacuum polarization by solitons},}\ }\href {\doibase
  10.1103/PhysRevD.30.2194} {\bibfield  {journal} {\bibinfo  {journal} {Phys.
  Rev. D}\ }\textbf {\bibinfo {volume} {30}},\ \bibinfo {pages} {2194--2200}
  (\bibinfo {year} {1984}{\natexlab{a}})}\BibitemShut {NoStop}%
\bibitem [{\citenamefont {Witten}(1979)}]{witten79}%
  \BibitemOpen
  \bibfield  {author} {\bibinfo {author} {\bibfnamefont {E.}~\bibnamefont
  {Witten}},\ }\bibfield  {title} {\enquote {\bibinfo {title} {Dyons of charge
  e$\theta / 2\pi$},}\ }\href {\doibase 10.1016/0370-2693(79)90838-4}
  {\bibfield  {journal} {\bibinfo  {journal} {Phys. Lett. B}\ }\textbf
  {\bibinfo {volume} {86}},\ \bibinfo {pages} {283--287} (\bibinfo {year}
  {1979})}\BibitemShut {NoStop}%
\bibitem [{\citenamefont {MacKenzie}\ and\ \citenamefont
  {Wilczek}(1984{\natexlab{b}})}]{macwilczek84}%
  \BibitemOpen
  \bibfield  {author} {\bibinfo {author} {\bibfnamefont {R.}~\bibnamefont
  {MacKenzie}}\ and\ \bibinfo {author} {\bibfnamefont {Frank}\ \bibnamefont
  {Wilczek}},\ }\bibfield  {title} {\enquote {\bibinfo {title} {Examples of
  vacuum polarization by solitons},}\ }\href {\doibase
  10.1103/PhysRevD.30.2260} {\bibfield  {journal} {\bibinfo  {journal} {Phys.
  Rev. D}\ }\textbf {\bibinfo {volume} {30}},\ \bibinfo {pages} {2260--2263}
  (\bibinfo {year} {1984}{\natexlab{b}})}\BibitemShut {NoStop}%
\bibitem [{\citenamefont {Gousheh}\ and\ \citenamefont
  {Lopez-Mobilia}(1994)}]{gousheh94}%
  \BibitemOpen
  \bibfield  {author} {\bibinfo {author} {\bibfnamefont {S.~S.}\ \bibnamefont
  {Gousheh}}\ and\ \bibinfo {author} {\bibfnamefont {R.}~\bibnamefont
  {Lopez-Mobilia}},\ }\bibfield  {title} {\enquote {\bibinfo {title} {Vacuum
  polarization by solitons in (1+1) dimensions},}\ }\href {\doibase
  10.1016/0550-3213(94)90198-8} {\bibfield  {journal} {\bibinfo  {journal}
  {Nuclear Physics B}\ }\textbf {\bibinfo {volume} {428}},\ \bibinfo {pages}
  {189--208} (\bibinfo {year} {1994})}\BibitemShut {NoStop}%
\bibitem [{\citenamefont {Charmchi}\ and\ \citenamefont
  {Gousheh}(2014)}]{charmchi14}%
  \BibitemOpen
  \bibfield  {author} {\bibinfo {author} {\bibfnamefont {F.}~\bibnamefont
  {Charmchi}}\ and\ \bibinfo {author} {\bibfnamefont {S.~S.}\ \bibnamefont
  {Gousheh}},\ }\bibfield  {title} {\enquote {\bibinfo {title} {Massive
  $\text{J}$ackiw - $\text{R}$ebbi model},}\ }\href {\doibase
  10.1016/j.nuclphysb.2014.03.021} {\bibfield  {journal} {\bibinfo  {journal}
  {Nucl. Phys. B}\ }\textbf {\bibinfo {volume} {883}},\ \bibinfo {pages}
  {256--266} (\bibinfo {year} {2014})}\BibitemShut {NoStop}%
\bibitem [{\citenamefont {Su}\ \emph {et~al.}(1979)\citenamefont {Su},
  \citenamefont {Schrieffer},\ and\ \citenamefont {Heeger}}]{su79}%
  \BibitemOpen
  \bibfield  {author} {\bibinfo {author} {\bibfnamefont {W.~P.}\ \bibnamefont
  {Su}}, \bibinfo {author} {\bibfnamefont {J.~R.}\ \bibnamefont {Schrieffer}},
  \ and\ \bibinfo {author} {\bibfnamefont {A.~J.}\ \bibnamefont {Heeger}},\
  }\bibfield  {title} {\enquote {\bibinfo {title} {Solitons in
  polyacetylene},}\ }\href {\doibase 10.1103/PhysRevLett.42.1698} {\bibfield
  {journal} {\bibinfo  {journal} {Phys. Rev. Lett.}\ }\textbf {\bibinfo
  {volume} {42}},\ \bibinfo {pages} {1698--1701} (\bibinfo {year}
  {1979})}\BibitemShut {NoStop}%
\bibitem [{\citenamefont {Su}\ and\ \citenamefont {Schrieffer}(1981)}]{su81}%
  \BibitemOpen
  \bibfield  {author} {\bibinfo {author} {\bibfnamefont {W.~P.}\ \bibnamefont
  {Su}}\ and\ \bibinfo {author} {\bibfnamefont {J.~R.}\ \bibnamefont
  {Schrieffer}},\ }\bibfield  {title} {\enquote {\bibinfo {title} {Fractionally
  charged excitations in charge-density-wave systems with commensurability
  3},}\ }\href {\doibase 10.1103/PhysRevLett.46.738} {\bibfield  {journal}
  {\bibinfo  {journal} {Phys. Rev. Lett.}\ }\textbf {\bibinfo {volume} {46}},\
  \bibinfo {pages} {738--741} (\bibinfo {year} {1981})}\BibitemShut {NoStop}%
\bibitem [{\citenamefont {Niemi}\ and\ \citenamefont
  {Semenoff}(1986)}]{semen86}%
  \BibitemOpen
  \bibfield  {author} {\bibinfo {author} {\bibfnamefont {A.~J.}\ \bibnamefont
  {Niemi}}\ and\ \bibinfo {author} {\bibfnamefont {G.~W.}\ \bibnamefont
  {Semenoff}},\ }\bibfield  {title} {\enquote {\bibinfo {title} {Fermion number
  fractionization in quantum field theory},}\ }\href {\doibase
  10.1016/0370-1573(86)90167-5} {\bibfield  {journal} {\bibinfo  {journal}
  {Phys. Rep.}\ }\textbf {\bibinfo {volume} {135}},\ \bibinfo {pages} {99--193}
  (\bibinfo {year} {1986})}\BibitemShut {NoStop}%
\bibitem [{\citenamefont {Rice}\ and\ \citenamefont {Mele}(1982)}]{rice82}%
  \BibitemOpen
  \bibfield  {author} {\bibinfo {author} {\bibfnamefont {M.~J.}\ \bibnamefont
  {Rice}}\ and\ \bibinfo {author} {\bibfnamefont {E.~J.}\ \bibnamefont
  {Mele}},\ }\bibfield  {title} {\enquote {\bibinfo {title} {Elementary
  excitations of a linearly conjugated diatomic polymer},}\ }\href {\doibase
  10.1103/PhysRevLett.49.1455} {\bibfield  {journal} {\bibinfo  {journal}
  {Phys. Rev. Lett.}\ }\textbf {\bibinfo {volume} {49}},\ \bibinfo {pages}
  {1455--1459} (\bibinfo {year} {1982})}\BibitemShut {NoStop}%
\bibitem [{\citenamefont {Jackiw}\ and\ \citenamefont
  {Semenoff}(1983)}]{jacksemen83}%
  \BibitemOpen
  \bibfield  {author} {\bibinfo {author} {\bibfnamefont {R.}~\bibnamefont
  {Jackiw}}\ and\ \bibinfo {author} {\bibfnamefont {G.}~\bibnamefont
  {Semenoff}},\ }\bibfield  {title} {\enquote {\bibinfo {title} {Continuum
  quantum field theory for a linearly conjugated diatomic polymer with fermion
  fractionization},}\ }\href {\doibase 10.1103/PhysRevLett.50.439} {\bibfield
  {journal} {\bibinfo  {journal} {Phys. Rev. Lett.}\ }\textbf {\bibinfo
  {volume} {50}},\ \bibinfo {pages} {439--442} (\bibinfo {year}
  {1983})}\BibitemShut {NoStop}%
\bibitem [{\citenamefont {Heeger}\ \emph {et~al.}(1988)\citenamefont {Heeger},
  \citenamefont {Kivelson}, \citenamefont {Schrieffer},\ and\ \citenamefont
  {Su}}]{heeger88}%
  \BibitemOpen
  \bibfield  {author} {\bibinfo {author} {\bibfnamefont {A.~J.}\ \bibnamefont
  {Heeger}}, \bibinfo {author} {\bibfnamefont {S.}~\bibnamefont {Kivelson}},
  \bibinfo {author} {\bibfnamefont {J.~R.}\ \bibnamefont {Schrieffer}}, \ and\
  \bibinfo {author} {\bibfnamefont {W.~P.}\ \bibnamefont {Su}},\ }\bibfield
  {title} {\enquote {\bibinfo {title} {Solitons in conducting polymers},}\
  }\href {\doibase 10.1103/RevModPhys.60.781} {\bibfield  {journal} {\bibinfo
  {journal} {Rev. Mod. Phys.}\ }\textbf {\bibinfo {volume} {60}},\ \bibinfo
  {pages} {781--850} (\bibinfo {year} {1988})}\BibitemShut {NoStop}%
\bibitem [{\citenamefont {Steinberg}\ \emph {et~al.}(2008)\citenamefont
  {Steinberg}, \citenamefont {Barak},\ and\ \citenamefont
  {Yacoby}}]{steinberg08}%
  \BibitemOpen
  \bibfield  {author} {\bibinfo {author} {\bibfnamefont {H.}~\bibnamefont
  {Steinberg}}, \bibinfo {author} {\bibfnamefont {G.}~\bibnamefont {Barak}}, \
  and\ \bibinfo {author} {\bibfnamefont {A.}~\bibnamefont {Yacoby}},\
  }\bibfield  {title} {\enquote {\bibinfo {title} {Charge fractionalization in
  quantum wires},}\ }\href {\doibase 10.1038/nphys810} {\bibfield  {journal}
  {\bibinfo  {journal} {Nature Phys.}\ }\textbf {\bibinfo {volume} {4}},\
  \bibinfo {pages} {116--119} (\bibinfo {year} {2008})}\BibitemShut {NoStop}%
\bibitem [{\citenamefont {Maciejko}\ and\ \citenamefont
  {Fiete}(2015)}]{maciejko15}%
  \BibitemOpen
  \bibfield  {author} {\bibinfo {author} {\bibfnamefont {J.}~\bibnamefont
  {Maciejko}}\ and\ \bibinfo {author} {\bibfnamefont {G.}~\bibnamefont
  {Fiete}},\ }\bibfield  {title} {\enquote {\bibinfo {title} {Fractionalized
  topological insulators},}\ }\href {\doibase 10.1038/nphys3311} {\bibfield
  {journal} {\bibinfo  {journal} {Nature Phys.}\ }\textbf {\bibinfo {volume}
  {11}},\ \bibinfo {pages} {385--388} (\bibinfo {year} {2015})}\BibitemShut
  {NoStop}%
\bibitem [{\citenamefont {Stern}(2016)}]{stern16}%
  \BibitemOpen
  \bibfield  {author} {\bibinfo {author} {\bibfnamefont {A.}~\bibnamefont
  {Stern}},\ }\bibfield  {title} {\enquote {\bibinfo {title} {Fractional
  topological insulators: A pedagogical review},}\ }\href {\doibase
  10.1146/annurev-conmatphys-031115-011559} {\bibfield  {journal} {\bibinfo
  {journal} {Ann. Rev. Condens. Mat. Phys.}\ }\textbf {\bibinfo {volume} {7}},\
  \bibinfo {pages} {349--368} (\bibinfo {year} {2016})}\BibitemShut {NoStop}%
\bibitem [{\citenamefont {Manton}(1983)}]{manton83}%
  \BibitemOpen
  \bibfield  {author} {\bibinfo {author} {\bibfnamefont {N.~S.}\ \bibnamefont
  {Manton}},\ }\bibfield  {title} {\enquote {\bibinfo {title} {Topology in the
  \text{W}einberg-\text{S}alam theory},}\ }\href {\doibase
  10.1103/PhysRevD.28.2019} {\bibfield  {journal} {\bibinfo  {journal} {Phys.
  Rev. D}\ }\textbf {\bibinfo {volume} {28}},\ \bibinfo {pages} {2019--2026}
  (\bibinfo {year} {1983})}\BibitemShut {NoStop}%
\bibitem [{\citenamefont {Manton}\ and\ \citenamefont
  {Sutcliffe}(2004)}]{manton04}%
  \BibitemOpen
  \bibfield  {author} {\bibinfo {author} {\bibfnamefont {N.~S.}\ \bibnamefont
  {Manton}}\ and\ \bibinfo {author} {\bibfnamefont {P.}~\bibnamefont
  {Sutcliffe}},\ }\bibfield  {title} {\enquote {\bibinfo {title} {Saddle points
  - sphalerons, in topological solitons},}\ }\href {\doibase
  10.1017/CBO9780511617034.012} {\bibfield  {journal} {\bibinfo  {journal}
  {Cambridge Monographs on Mathematical Physics}\ ,\ \bibinfo {pages}
  {441--466}} (\bibinfo {year} {2004})}\BibitemShut {NoStop}%
\bibitem [{\citenamefont {Klinkhamer}\ and\ \citenamefont
  {Rupp}(2003)}]{klink03}%
  \BibitemOpen
  \bibfield  {author} {\bibinfo {author} {\bibfnamefont {F.~R.}\ \bibnamefont
  {Klinkhamer}}\ and\ \bibinfo {author} {\bibfnamefont {C.}~\bibnamefont
  {Rupp}},\ }\bibfield  {title} {\enquote {\bibinfo {title} {Sphalerons,
  spectral flow and anomalies},}\ }\href {\doibase 10.1063/1.1590420}
  {\bibfield  {journal} {\bibinfo  {journal} {Jour. of Math. Phys.}\ }\textbf
  {\bibinfo {volume} {44}},\ \bibinfo {pages} {3619} (\bibinfo {year}
  {2003})}\BibitemShut {NoStop}%
\bibitem [{\citenamefont {Klinkhamer}\ and\ \citenamefont
  {Manton}(1984)}]{klink84}%
  \BibitemOpen
  \bibfield  {author} {\bibinfo {author} {\bibfnamefont {F.~R.}\ \bibnamefont
  {Klinkhamer}}\ and\ \bibinfo {author} {\bibfnamefont {N.~S.}\ \bibnamefont
  {Manton}},\ }\bibfield  {title} {\enquote {\bibinfo {title} {A saddle - point
  solution in the \text{W}einberg-\text{S}alam theory},}\ }\href {\doibase
  10.1103/PhysRevD.30.2212} {\bibfield  {journal} {\bibinfo  {journal} {Phys.
  Rev. D}\ }\textbf {\bibinfo {volume} {30}},\ \bibinfo {pages} {2212--2220}
  (\bibinfo {year} {1984})}\BibitemShut {NoStop}%
\bibitem [{\citenamefont {Rubakov}\ and\ \citenamefont
  {Shaposhnikov}(1996)}]{rubakov96}%
  \BibitemOpen
  \bibfield  {author} {\bibinfo {author} {\bibfnamefont {V.~A.}\ \bibnamefont
  {Rubakov}}\ and\ \bibinfo {author} {\bibfnamefont {M.~E.}\ \bibnamefont
  {Shaposhnikov}},\ }\bibfield  {title} {\enquote {\bibinfo {title}
  {Electroweak baryon number non-conservation in the early \text{U}niverse and
  in high - energy collisions},}\ }\href {\doibase
  10.1070/PU1996v039n05ABEH000145} {\bibfield  {journal} {\bibinfo  {journal}
  {Physics - Uspekhi}\ }\textbf {\bibinfo {volume} {39}} (\bibinfo {year}
  {1996}),\ 10.1070/PU1996v039n05ABEH000145}\BibitemShut {NoStop}%
\bibitem [{\citenamefont {Rostam~Zadeh}\ and\ \citenamefont
  {Gousheh}(2019)}]{shiva18}%
  \BibitemOpen
  \bibfield  {author} {\bibinfo {author} {\bibfnamefont {S.}~\bibnamefont
  {Rostam~Zadeh}}\ and\ \bibinfo {author} {\bibfnamefont {S.~S.}\ \bibnamefont
  {Gousheh}},\ }\bibfield  {title} {\enquote {\bibinfo {title} {Minimal system
  including weak sphalerons for investigating the evolution of matter
  asymmetries and hypermagnetic fields},}\ }\href {\doibase
  10.1103/PhysRevD.99.096009} {\bibfield  {journal} {\bibinfo  {journal} {Phys.
  Rev. D}\ }\textbf {\bibinfo {volume} {99}},\ \bibinfo {pages} {096009}
  (\bibinfo {year} {2019})}\BibitemShut {NoStop}%
\bibitem [{\citenamefont {Dashen}\ \emph {et~al.}(1974)\citenamefont {Dashen},
  \citenamefont {Hasslacher},\ and\ \citenamefont {Neveu}}]{dhn74}%
  \BibitemOpen
  \bibfield  {author} {\bibinfo {author} {\bibfnamefont {Roger~F.}\
  \bibnamefont {Dashen}}, \bibinfo {author} {\bibfnamefont {Brosl}\
  \bibnamefont {Hasslacher}}, \ and\ \bibinfo {author} {\bibfnamefont
  {Andr\'e}\ \bibnamefont {Neveu}},\ }\bibfield  {title} {\enquote {\bibinfo
  {title} {Nonperturbative methods and extended-hadron models in field theory.
  iii. four-dimensional non-\text{A}belian models},}\ }\href {\doibase
  10.1103/PhysRevD.10.4138} {\bibfield  {journal} {\bibinfo  {journal} {Phys.
  Rev. D}\ }\textbf {\bibinfo {volume} {10}},\ \bibinfo {pages} {4138}
  (\bibinfo {year} {1974})}\BibitemShut {NoStop}%
\bibitem [{\citenamefont {Boguta}(1983)}]{boguta83}%
  \BibitemOpen
  \bibfield  {author} {\bibinfo {author} {\bibfnamefont {J.}~\bibnamefont
  {Boguta}},\ }\bibfield  {title} {\enquote {\bibinfo {title} {Can nuclear
  interactions be long ranged?}}\ }\href {\doibase 10.1103/PhysRevLett.50.148}
  {\bibfield  {journal} {\bibinfo  {journal} {Phys. Rev. Lett.}\ }\textbf
  {\bibinfo {volume} {50}},\ \bibinfo {pages} {148--152} (\bibinfo {year}
  {1983})}\BibitemShut {NoStop}%
\bibitem [{\citenamefont {Nohl}(1975)}]{nohl75}%
  \BibitemOpen
  \bibfield  {author} {\bibinfo {author} {\bibfnamefont {Craig~R.}\
  \bibnamefont {Nohl}},\ }\bibfield  {title} {\enquote {\bibinfo {title}
  {Bound-state solutions of the \text{D}irac equation in extended hadron
  models},}\ }\href {\doibase 10.1103/PhysRevD.12.1840} {\bibfield  {journal}
  {\bibinfo  {journal} {Phys. Rev. D}\ }\textbf {\bibinfo {volume} {12}},\
  \bibinfo {pages} {1840--1842} (\bibinfo {year} {1975})}\BibitemShut {NoStop}%
\bibitem [{\citenamefont {Boguta}\ and\ \citenamefont {Kunz}(1985)}]{boguta85}%
  \BibitemOpen
  \bibfield  {author} {\bibinfo {author} {\bibfnamefont {J.}~\bibnamefont
  {Boguta}}\ and\ \bibinfo {author} {\bibfnamefont {J.}~\bibnamefont {Kunz}},\
  }\bibfield  {title} {\enquote {\bibinfo {title} {Hadroids and sphalerons},}\
  }\href {\doibase 10.1016/0370-2693(85)90419-8} {\bibfield  {journal}
  {\bibinfo  {journal} {Phys. Lett. B}\ }\textbf {\bibinfo {volume} {154}},\
  \bibinfo {pages} {407--410} (\bibinfo {year} {1985})}\BibitemShut {NoStop}%
\bibitem [{\citenamefont {Ringwald}(1988)}]{ringwald88}%
  \BibitemOpen
  \bibfield  {author} {\bibinfo {author} {\bibfnamefont {A.}~\bibnamefont
  {Ringwald}},\ }\bibfield  {title} {\enquote {\bibinfo {title} {Sphaleron and
  level crossing},}\ }\href {\doibase 10.1016/0370-2693(88)91047-7} {\bibfield
  {journal} {\bibinfo  {journal} {Phys. Lett. B}\ }\textbf {\bibinfo {volume}
  {213}},\ \bibinfo {pages} {61--63} (\bibinfo {year} {1988})}\BibitemShut
  {NoStop}%
\bibitem [{\citenamefont {Jutta~Kunz}(1993)}]{kunz93}%
  \BibitemOpen
  \bibfield  {author} {\bibinfo {author} {\bibfnamefont {Yves Brihaye~and}\
  \bibnamefont {Jutta~Kunz}},\ }\bibfield  {title} {\enquote {\bibinfo {title}
  {Fermions in the background of the sphaleron barrier},}\ }\href {\doibase
  10.1016/0370-2693(93)91413-H} {\bibfield  {journal} {\bibinfo  {journal}
  {Phys. Lett. B}\ }\textbf {\bibinfo {volume} {304}},\ \bibinfo {pages}
  {141--146} (\bibinfo {year} {1993})}\BibitemShut {NoStop}%
\bibitem [{\citenamefont {Yaffe}(1989)}]{yaffe89}%
  \BibitemOpen
  \bibfield  {author} {\bibinfo {author} {\bibfnamefont {Laurence~G.}\
  \bibnamefont {Yaffe}},\ }\bibfield  {title} {\enquote {\bibinfo {title}
  {Static solutions of \text{SU}(2)-\text{H}iggs theory},}\ }\href {\doibase
  10.1103/PhysRevD.40.3463} {\bibfield  {journal} {\bibinfo  {journal} {Phys.
  Rev. D}\ }\textbf {\bibinfo {volume} {40}},\ \bibinfo {pages} {3463--3473}
  (\bibinfo {year} {1989})}\BibitemShut {NoStop}%
\bibitem [{\citenamefont {Brihaye}\ and\ \citenamefont {Kunz}(1994)}]{kunz94}%
  \BibitemOpen
  \bibfield  {author} {\bibinfo {author} {\bibfnamefont {Yves}\ \bibnamefont
  {Brihaye}}\ and\ \bibinfo {author} {\bibfnamefont {Jutta}\ \bibnamefont
  {Kunz}},\ }\bibfield  {title} {\enquote {\bibinfo {title} {Axially symmetric
  solutions in electroweak theory},}\ }\href {\doibase
  10.1103/PhysRevD.50.4175} {\bibfield  {journal} {\bibinfo  {journal} {Phys.
  Rev. D}\ }\textbf {\bibinfo {volume} {50}},\ \bibinfo {pages} {4175--4182}
  (\bibinfo {year} {1994})}\BibitemShut {NoStop}%
\bibitem [{\citenamefont {Nolte}\ \emph {et~al.}(1996)\citenamefont {Nolte},
  \citenamefont {Kunz},\ and\ \citenamefont {Kleihaus}}]{nolte96}%
  \BibitemOpen
  \bibfield  {author} {\bibinfo {author} {\bibfnamefont {Guido}\ \bibnamefont
  {Nolte}}, \bibinfo {author} {\bibfnamefont {Jutta}\ \bibnamefont {Kunz}}, \
  and\ \bibinfo {author} {\bibfnamefont {Burkhard}\ \bibnamefont {Kleihaus}},\
  }\bibfield  {title} {\enquote {\bibinfo {title} {Nondegenerate fermions in
  the background of the sphaleron barrier},}\ }\href {\doibase
  10.1103/PhysRevD.53.3451} {\bibfield  {journal} {\bibinfo  {journal} {Phys.
  Rev. D}\ }\textbf {\bibinfo {volume} {53}},\ \bibinfo {pages} {3451--3459}
  (\bibinfo {year} {1996})}\BibitemShut {NoStop}%
\bibitem [{\citenamefont {Klinkhamer}\ and\ \citenamefont
  {Manton}(1985)}]{klink84b}%
  \BibitemOpen
  \bibfield  {author} {\bibinfo {author} {\bibfnamefont {F.~R.}\ \bibnamefont
  {Klinkhamer}}\ and\ \bibinfo {author} {\bibfnamefont {N.~S.}\ \bibnamefont
  {Manton}},\ }\bibfield  {title} {\enquote {\bibinfo {title} {A saddle-point
  solution in the \text{W}einberg-\text{S}alam theory},}\ }\href
  {http://inspirehep.net/record/1610374/files/chap%209%20a%20saddle%20point%20solution.pdf}
  {\bibfield  {journal} {\bibinfo  {journal} {Design and Utilization of the
  Superconducting Super Collider, Proceedings 1984 Summer Study, Snowmass, USA,
  Editors R. Donaldson and J.G. Morfin, New York, USA}\ ,\ \bibinfo {pages}
  {805--806}} (\bibinfo {year} {1985})}\BibitemShut {NoStop}%
\bibitem [{\citenamefont {Klinkhamer}(1990)}]{klink90}%
  \BibitemOpen
  \bibfield  {author} {\bibinfo {author} {\bibfnamefont {F.~R.}\ \bibnamefont
  {Klinkhamer}},\ }\bibfield  {title} {\enquote {\bibinfo {title} {A new
  sphaleron in the \text{W}einberg-\text{S}alam theory},}\ }\href {\doibase
  10.1016/0370-2693(90)91319-7} {\bibfield  {journal} {\bibinfo  {journal}
  {Phys. Lett. B}\ }\textbf {\bibinfo {volume} {246}},\ \bibinfo {pages}
  {131--134} (\bibinfo {year} {1990})}\BibitemShut {NoStop}%
\bibitem [{\citenamefont {Witten}(1977)}]{witten77}%
  \BibitemOpen
  \bibfield  {author} {\bibinfo {author} {\bibfnamefont {Edward}\ \bibnamefont
  {Witten}},\ }\bibfield  {title} {\enquote {\bibinfo {title} {Some exact
  multipseudoparticle solutions of classical \text{Y}ang-\text{M}ills
  theory},}\ }\href {\doibase 10.1103/PhysRevLett.38.121} {\bibfield  {journal}
  {\bibinfo  {journal} {Phys. Rev. Lett.}\ }\textbf {\bibinfo {volume} {38}},\
  \bibinfo {pages} {121--124} (\bibinfo {year} {1977})}\BibitemShut {NoStop}%
\bibitem [{\citenamefont {Ibadov}\ \emph {et~al.}(2008)\citenamefont {Ibadov},
  \citenamefont {Kleihaus}, \citenamefont {Kunz},\ and\ \citenamefont
  {Leißner}}]{kunz08}%
  \BibitemOpen
  \bibfield  {author} {\bibinfo {author} {\bibfnamefont {Rostam}\ \bibnamefont
  {Ibadov}}, \bibinfo {author} {\bibfnamefont {Burkhard}\ \bibnamefont
  {Kleihaus}}, \bibinfo {author} {\bibfnamefont {Jutta}\ \bibnamefont {Kunz}},
  \ and\ \bibinfo {author} {\bibfnamefont {Michael}\ \bibnamefont {Leißner}},\
  }\bibfield  {title} {\enquote {\bibinfo {title} {Gravitating sphaleron -
  antisphaleron systems},}\ }\href {\doibase 10.1016/j.physletb.2008.03.055}
  {\bibfield  {journal} {\bibinfo  {journal} {Phys. Lett. B}\ }\textbf
  {\bibinfo {volume} {663}},\ \bibinfo {pages} {136--140} (\bibinfo {year}
  {2008})}\BibitemShut {NoStop}%
\bibitem [{\citenamefont {Kleihaus}\ \emph {et~al.}(2008)\citenamefont
  {Kleihaus}, \citenamefont {Kunz},\ and\ \citenamefont {Leißner}}]{kunz08b}%
  \BibitemOpen
  \bibfield  {author} {\bibinfo {author} {\bibfnamefont {Burkhard}\
  \bibnamefont {Kleihaus}}, \bibinfo {author} {\bibfnamefont {Jutta}\
  \bibnamefont {Kunz}}, \ and\ \bibinfo {author} {\bibfnamefont {Michael}\
  \bibnamefont {Leißner}},\ }\bibfield  {title} {\enquote {\bibinfo {title}
  {Sphalerons, antisphalerons and vortex rings},}\ }\href {\doibase
  10.1016/j.physletb.2008.04.027} {\bibfield  {journal} {\bibinfo  {journal}
  {Phys. Lett. B}\ }\textbf {\bibinfo {volume} {663}},\ \bibinfo {pages}
  {438--444} (\bibinfo {year} {2008})}\BibitemShut {NoStop}%
\bibitem [{\citenamefont {Kleihaus}\ \emph {et~al.}(2004)\citenamefont
  {Kleihaus}, \citenamefont {Kunz},\ and\ \citenamefont {Myklevoll}}]{kunz04}%
  \BibitemOpen
  \bibfield  {author} {\bibinfo {author} {\bibfnamefont {Burkhard}\
  \bibnamefont {Kleihaus}}, \bibinfo {author} {\bibfnamefont {Jutta}\
  \bibnamefont {Kunz}}, \ and\ \bibinfo {author} {\bibfnamefont {Kari}\
  \bibnamefont {Myklevoll}},\ }\bibfield  {title} {\enquote {\bibinfo {title}
  {Platonic sphalerons},}\ }\href {\doibase 10.1016/j.physletb.2003.12.036}
  {\bibfield  {journal} {\bibinfo  {journal} {Phys. Lett. B}\ }\textbf
  {\bibinfo {volume} {582}},\ \bibinfo {pages} {187--195} (\bibinfo {year}
  {2004})}\BibitemShut {NoStop}%
\bibitem [{\citenamefont {Kleihaus}\ \emph {et~al.}(1999)\citenamefont
  {Kleihaus}, \citenamefont {Tchrakian},\ and\ \citenamefont
  {Zimmerschied}}]{kleihaus99}%
  \BibitemOpen
  \bibfield  {author} {\bibinfo {author} {\bibfnamefont {B.}~\bibnamefont
  {Kleihaus}}, \bibinfo {author} {\bibfnamefont {D.~H.}\ \bibnamefont
  {Tchrakian}}, \ and\ \bibinfo {author} {\bibfnamefont {F.}~\bibnamefont
  {Zimmerschied}},\ }\bibfield  {title} {\enquote {\bibinfo {title} {Sphaleron
  of a 4 dimensional \text{SO}(4) \text{H}iggs model},}\ }\href {\doibase
  10.1016/S0370-2693(99)00838-2} {\bibfield  {journal} {\bibinfo  {journal}
  {Phys. Lett. B}\ }\textbf {\bibinfo {volume} {461}},\ \bibinfo {pages}
  {224--229} (\bibinfo {year} {1999})}\BibitemShut {NoStop}%
\bibitem [{\citenamefont {Kleihaus}\ and\ \citenamefont
  {Kunz}(1994)}]{kunzmulti94}%
  \BibitemOpen
  \bibfield  {author} {\bibinfo {author} {\bibfnamefont {Burkhard}\
  \bibnamefont {Kleihaus}}\ and\ \bibinfo {author} {\bibfnamefont {Jutta}\
  \bibnamefont {Kunz}},\ }\bibfield  {title} {\enquote {\bibinfo {title}
  {Multisphalerons in the \text{W}einberg-\text{S}alam theory},}\ }\href
  {\doibase 10.1103/PhysRevD.50.5343} {\bibfield  {journal} {\bibinfo
  {journal} {Phys. Rev. D}\ }\textbf {\bibinfo {volume} {50}},\ \bibinfo
  {pages} {5343--5351} (\bibinfo {year} {1994})}\BibitemShut {NoStop}%
\bibitem [{\citenamefont {Klinkhamer}(1993)}]{klink93}%
  \BibitemOpen
  \bibfield  {author} {\bibinfo {author} {\bibfnamefont {F.~R.}\ \bibnamefont
  {Klinkhamer}},\ }\bibfield  {title} {\enquote {\bibinfo {title} {Construction
  of a new electroweak sphaleron},}\ }\href {\doibase
  10.1016/0550-3213(93)90437-T} {\bibfield  {journal} {\bibinfo  {journal}
  {Nucl. Phys. B}\ }\textbf {\bibinfo {volume} {410}},\ \bibinfo {pages}
  {343--354} (\bibinfo {year} {1993})}\BibitemShut {NoStop}%
\bibitem [{\citenamefont {Klinkhamer}\ and\ \citenamefont
  {Rupp}(2005)}]{klink05}%
  \BibitemOpen
  \bibfield  {author} {\bibinfo {author} {\bibfnamefont {F.~R.}\ \bibnamefont
  {Klinkhamer}}\ and\ \bibinfo {author} {\bibfnamefont {C.}~\bibnamefont
  {Rupp}},\ }\bibfield  {title} {\enquote {\bibinfo {title} {A sphaleron for
  the non-\text{A}belian anomaly},}\ }\href {\doibase
  10.1016/j.nuclphysb.2004.12.027} {\bibfield  {journal} {\bibinfo  {journal}
  {Nucl. Phys. B}\ }\textbf {\bibinfo {volume} {709}},\ \bibinfo {pages}
  {171--191} (\bibinfo {year} {2005})}\BibitemShut {NoStop}%
\bibitem [{\citenamefont {Klinkhamer}\ and\ \citenamefont
  {Nagel}(2017)}]{klink17}%
  \BibitemOpen
  \bibfield  {author} {\bibinfo {author} {\bibfnamefont {F.~R.}\ \bibnamefont
  {Klinkhamer}}\ and\ \bibinfo {author} {\bibfnamefont {P.}~\bibnamefont
  {Nagel}},\ }\bibfield  {title} {\enquote {\bibinfo {title} {\text{SU}(3)
  sphaleron: Numerical solution},}\ }\href {\doibase
  10.1103/PhysRevD.96.016006} {\bibfield  {journal} {\bibinfo  {journal} {Phys.
  Rev. D}\ }\textbf {\bibinfo {volume} {96}},\ \bibinfo {pages} {016006}
  (\bibinfo {year} {2017})}\BibitemShut {NoStop}%
\bibitem [{\citenamefont {Kleihaus}\ \emph {et~al.}(1991)\citenamefont
  {Kleihaus}, \citenamefont {Kunz},\ and\ \citenamefont {Brihaye}}]{kunz91}%
  \BibitemOpen
  \bibfield  {author} {\bibinfo {author} {\bibfnamefont {B.}~\bibnamefont
  {Kleihaus}}, \bibinfo {author} {\bibfnamefont {J.}~\bibnamefont {Kunz}}, \
  and\ \bibinfo {author} {\bibfnamefont {Y.}~\bibnamefont {Brihaye}},\
  }\bibfield  {title} {\enquote {\bibinfo {title} {The electroweak sphaleron at
  physical mixing angle},}\ }\href {\doibase 10.1016/0370-2693(91)90560-D}
  {\bibfield  {journal} {\bibinfo  {journal} {Phys. Lett. B}\ }\textbf
  {\bibinfo {volume} {273}},\ \bibinfo {pages} {100--104} (\bibinfo {year}
  {1991})}\BibitemShut {NoStop}%
\end{thebibliography}%

\end{document}